%% file: mri.tex
\documentclass{aa}
\usepackage{amsmath,graphicx,amssymb}
\usepackage{txfonts}
\usepackage{natbib}
\bibpunct{(}{)}{,}{a}{}{,}

\newcommand{\zav}[1]{\left(#1\right)}
\newcommand\req{\ensuremath{R_\text{eq}}}
\newcommand{\hzav}[1]{\left[#1\right]}
\newcommand{\azav}[1]{\left|#1\right|}
\newcommand{\szav}[1]{\left\{#1\right\}}

\newlength\staretab
\newcommand{\Teff}{\mbox{$T_\mathrm{eff}$}}
\newcommand{\de}{\text{d}}
\newcommand\x[1]{\ensuremath{#1_\text{X}}}

\makeatletter
\def\sgn{\mathop{\operator@font sgn}\nolimits}
\makeatother

\begin{document}

\title{Magnetorotational instability in decretion
disks of critically rotating stars and the outer structure of Be and Be/X-ray
disks}

\author{J.~Krti\v{c}ka \and P. Kurf\"urst \and I.~Krti\v{c}kov\'a}


\institute{Department of Theoretical Physics and Astrophysics,
           Masaryk University, Kotl\'a\v rsk\' a 2, CZ-611\,37 Brno, Czech
           Republic}

\date{Received}

\abstract {Evolutionary models of fast-rotating stars show that the stellar
rotational velocity may approach the critical speed. Critically rotating stars
cannot spin up more, therefore they lose their excess angular momentum through
an equatorial outflowing disk. The radial extension of such disks is unknown,
partly because we lack information about the radial variations of the
viscosity.}{We study the magnetorotational instability, which is considered to
be the origin of anomalous viscosity in outflowing disks.}{We used analytic
calculations to study the stability of outflowing disks submerged in the
magnetic field.}{The magnetorotational instability develops close to the star if
the plasma parameter is large enough. At large radii the instability disappears
in the region where the disk orbital velocity is roughly equal to the sound
speed.}{The magnetorotational instability is a plausible source of anomalous
viscosity in outflowing disks. This is also true in the region where the disk radial
velocity approaches the sound speed. The disk sonic radius can therefore be 
roughly considered as an effective outer disk radius, although disk material
may escape from the star to the insterstellar medium. The radial profile of the
angular momentum-loss rate already flattens there, consequently, the disk mass-loss rate can be calculated with the sonic radius as the effective disk outer
radius. We discuss a possible observation determination of the outer disk radius
by using Be and Be/X-ray binaries.}


\titlerunning{Magnetorotational instability in the
disks of critically rotating stars}

\authorrunning{J.~Krti\v{c}ka, P. Kurf\"urst, and I.~Krti\v{c}kov\'a}
\maketitle

\section{Introduction}

Accretion is one of the most ubiquitous processes in astrophysics and occurs
in different environments on various length scales. The crucial problem of 
accretion is that the accreting matter has to lose its angular momentum on
Keplerian orbits while moving towards the centre of gravity. If the corresponding
angular momentum transfer would have to rely on the molecular viscosity alone,
the time of accretion would be prohibitively long. Consequently, some type of
anomalous viscosity has to be present in disks that allows the accretion to
proceed in time compatible with observational constraints. Magnetorotational
instability \citep[MRI,][]{barbus} is considered to be the main source of
anomalous viscosity in such disks.

In contrast, decretion disks are connected with stellar mass-loss, but
their presence stems from the same principle as the existence of the accretion
disk, that is, the need for angular momentum transport. Evolutionary models of
fast-rotating stars show that the stellar rotational velocity may approach the
critical velocity \citep{mee,ekmemaba}, above which no stellar spin-up is
possible. The further evolution of a critically rotating star may require loss of
angular momentum, which is achieved through an outflowing decretion disk
\citep{kom}. The disk may extend to several hundred stellar radii
\citep{kurfek} or even more. The angular momentum transport in the disk
requires some type of anomalous viscosity, which is presumably connected with
MRI.

Decretion disks are typically connected with Be and B[e] star disks
\citep{los91,sapporo}. Classical Be stars are non-supergiant B stars, whose
hydrogen lines have (or had in the past) an emission component that can be
explained by an equatorial disk \citep[see][for a review]{ricam}. The
observational evidence supports the picture that the material of the disk comes
from the star, and therefore the disks are called decretion disks. The mechanism
that transports the stellar material into the disk is unclear at present.
However, it is generally believed that once the material appears in the disk, it
has a Keplerian rotation and is transported away from the star by the same
anomalous viscosity that operates in accretion disks.

However, it is unclear whether the MRI can operate in decretion disks of Be
stars. From an observational and theoretical point of view, the Be star disks
can extend to the distances of at least hundreds of stellar radii, and it is
unclear whether the angular momentum transport in the distant regions is the
same as close to the star. Therefore, we provide here an analytic study of MRI
in the decretion disks of Be stars, which will be accompanied by a detailed
magnetohydrodynamical modelling in a future paper.

\section{Basic physics of MRI}

Magnetorotational instability can exist even in very weakly magnetized disks. If
there is a radial displacement in such disks, the magnetic stresses may win and
push the material back towards its original position. These displacements do not
cause any instability. The displaced material is kept on corotation with respect
to its original position. If the centrifugal force acting on the displaced
material is higher than the force due to magnetic stresses, the material
accelerates in the direction of the original displacement, leading to an instability.

The behaviour of the displaced particle in the disk submerged in the magnetic field
with a zero radial component can be described by a dispersion relation
\citep{barbus}
\begin{equation}
\label{balhaw}
\frac{k_z^2+k_R^2}{k_z^2}\tilde\omega^4-
\hzav{\kappa^2+\zav{\frac{k_R}{k_z}N_z-N_R}^2}\tilde\omega^2-
4\Omega^2k_z^2 \varv_{\text{A}z}^2=0
\end{equation}
in the cylindrical coordinates $(R, \phi ,z)$, where
\begin{equation}
\tilde\omega^2=\omega^2-k_z^2 \varv_{\text{A}z}^2,
\end{equation}
the square of the epicyclic frequency is
\begin{equation}
\label{epicyc}
\kappa^2=\frac{2\Omega}{R}\frac{\de\zav{R^2\Omega}}{\de R},
\end{equation}
the pieces of the Brunt-V\"ais\"al\"a frequency $N^2=N_R^2+N_z^2$ are
\begin{subequations}
\label{jeseniusoba}
\begin{equation}
\label{jesenius}
N_R^2=-\frac{3}{5\rho}\frac{\partial P}{\partial R}
\frac{\partial\ln P\rho^{-5/3}}{\partial R},
\end{equation}
\begin{equation}
N_z^2=-\frac{3}{5\rho}\frac{\partial P}{\partial z}
\frac{\partial\ln P\rho^{-5/3}}{\partial z},
\end{equation}
\end{subequations}
$\Omega$ is the angular velocity, $k_z$ and $k_R$ are vertical and radial
wavenumbers, $\omega$ is the angular frequency, and the square of the
vertical Alfv\'en velocity is
\begin{equation}
\label{blansko}
\varv_{\text{A}z}^2=\frac{B_z^2}{4\pi\rho},
\end{equation}
where $B_z$ is the vertical component of the magnetic field and $\rho$ is the
density.
The instability occurs in the disks with radially decreasing angular velocity,
\begin{equation}
\label{lhota}
\frac{\de\Omega^2}{\de R}<0.
\end{equation}
In such disks the dispersion relation Eq.~\eqref{balhaw} leads to instability ($\omega^2<0$)
for vertical wavenumbers $k_z<k_{z,\text{max}}$ with
\begin{equation}
\label{zidenice}
k_{z,\text{max}}^2=\frac{1}{2\varv_{\text{A}z}^2}\szav{\hzav{
\zav{N^2+\frac{\de\Omega^2}{\de\ln R}}^2-
4N_z^2\frac{\de\Omega^2}{\de\ln R}}^{1/2}-
N^2-\frac{\de\Omega^2}{\de\ln R}}.
\end{equation}
This instability is thought to be the source of anomalous viscosity in accretion
disks.

\section{Stationary-disk model}
\label{stacdiskkap}

As a starting point of our instability study we assumed an outflowing disk
transporting an excess angular momentum from a critically rotating star
\citep{kom} with equatorial radius $R_\text{eq}=\frac{3}{2}R_*$ (here $R_*$ is
the stellar polar radius). The stationary-disk solution is derived from the
equation of continuity, and equations of motion in radial and azimuthal
directions. The static equations provide a radial dependence of the integrated
disk density $\Sigma$, and radial $\varv_R$ and azimuthal $\varv_\phi$
components of the velocity. For a simplicity, we assumed an isothermal disk with
$T=T_0\equiv\frac{1}{2}T_\text{eff}$, which roughly corresponds to the NLTE
models \citep{milma,carbjo}.

The stellar parameters selected for the modelling correspond to the
main-sequence B2--B3 star with an effective temperature
$\Teff=20\,000\,\text{K}$, mass $M=6.60\,M_\odot$, and radius
$R_*=3.71\,{R}_\odot$ \citep{har}. We also calculated additional models for
main-sequence stars with different effective temperatures
$\Teff=30\,000\,\text{K}$ and $\Teff=14\,000\,\text{K}$ and with radially
decreasing disk temperature $T=T_0(R_\text{eq}/R)^p$ with $p=0.1$ and $p=0.2$.
The results of our models show that the properties of the disk model and our
results in general do not significantly depend on the particular choice of the
disk and stellar parameters. Consequently, below we mainly 
discuss a model with $\Teff=20\,000\,\text{K}$.

\begin{figure}
\centering
\includegraphics[width=\hsize]{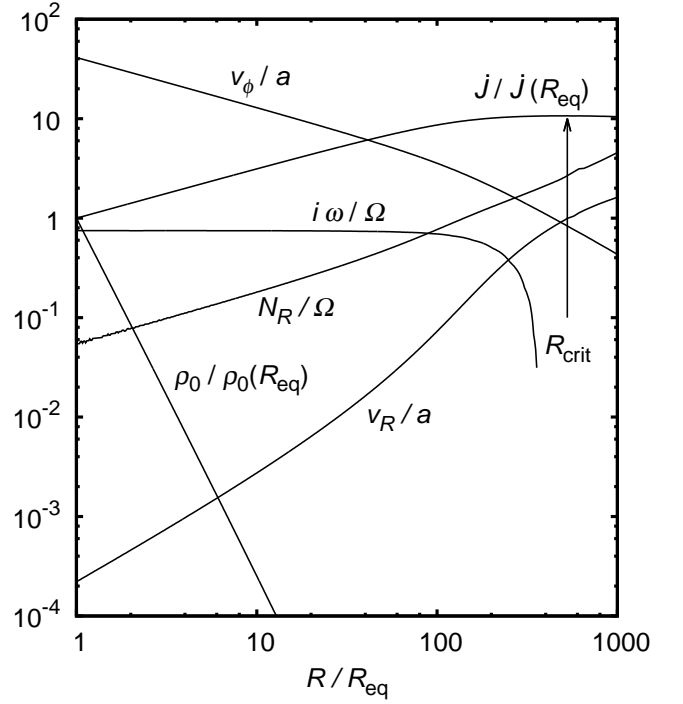}
\caption{Dependence of the radial and azimuthal velocities (in units of
sound speed $a$), the midplane disk density $\rho_0$ (relative to its value at
$R=R_\text{eq}$), the angular momentum-loss rate in units of equator release
angular momentum-loss rate $\dot J(\req)$, and the midplane maximum
MRI growth rate $i\omega$ and the radial part of the Brunt-V\"ais\"al\"a
frequency $N_R$ in units of the disk angular velocity on the radius in
a viscous disk. We assumed a viscosity parameter $\alpha=0.1$ here. The arrow
denotes the location of the critical point.}
\label{vd1uld}
\end{figure}

In Fig.~\ref{vd1uld} we provide the radial dependences of selected variables.
The radial disk velocity $\varv_R$ increases linearly
with radius close to the star. In the same region the azimuthal velocity follows
the Keplerian rotation $\varv_ \phi \sim R^{-1/2}$. The midplane disk density
$\rho_0$ strongly decreases with radius because of disk acceleration, flaring,
and geometrical reasons in total as $\rho_0\sim R^{-3.5}$. The angular
momentum loss $\dot J=R\varv_ \phi \dot M$ increases linearly with radius close
to the star and is highest close to the critical radius $R_\text{crit}$,
where the radial velocity reaches the sound speed, $\varv_R=a$
\citep{sapporo,kom}.

\section{MRI close to the star}
\label{nearstellar}

The condition of the decreasing angular velocity Eq.~\eqref{lhota} is fulfilled
everywhere in the Be star disks. However, the disks are typically very thin
close to the star. In a thin disk the lowest vertical wavenumber
Eq.~\eqref{zidenice} could be too low for the development of the instability.
Denoting a typical vertical disk thickness as $2^{3/2}H$ with
\begin{equation}
H=\frac{a}{\varv_\text{K}}R,
\end{equation}
where $\varv_\text{K}$ is the Keplerian rotation velocity at radius $R$ and $a$
is the sound speed, only the modes with wavelength lower than $2^{3/2}H$ can
exist in the disk, giving the condition for the lowest wavelength leading to
instability as
\begin{equation}
\label{instakh}
\frac{2\pi}{k_{z,\text{max}}}<2^{3/2}H.
\end{equation}
Close to the star the Brunt-V\"ais\"al\"a frequency
is negligible with respect to the angular velocity derivative in
Eq.~\eqref{zidenice}, yielding the wavenumbers leading to instability
\begin{equation}
\label{adamov}
k_z<k_{z,\text{max}}=\frac{1}{\varv_{\text{A}z}}
{\azav{\frac{\de\Omega^2}{\de\ln R}}}^{1/2}.
\end{equation}
Inserting the Alfv\'en speed Eq.~\eqref{blansko} with the highest density
corresponding to the midplane density $\rho_0$, the conditions for the
development of instability Eqs.~\eqref{instakh} and \eqref{adamov} can be
combined into a condition for the vertical magnetic field,
\begin{equation}
\label{bilovice}
B_z<\frac{\sqrt 6}{\pi}\zav{\sqrt{\frac{2}{\pi}}
\frac{a\varv_\text{K}\dot M}{\varv_RR^2}}^{1/2},
\end{equation}
where the disk mass-loss rate is $\dot M=\zav{2\pi}^{3/2}\varv_R\rho_0RH$
\citep[e.g.,][]{kom}, and we assumed a Keplerian rotation
$\Omega=\varv_\text{K}/R$. Physically, this means that the gas pressure has to
dominate magnetic pressure and the plasma parameter $\beta>\pi^2/3\approx3$
\citep{barbus,horal}. For lower values of $\beta$ only stable modes
($k_z>k_{z,\text{max}}$) can exist in the disk.

Eq.~\eqref{bilovice} can be rewritten in scaled quantities as
\begin{equation}
\label{bilovices}
B_z<250\,\text{G}\hzav{\zav{\frac{a/\varv_R}{10^{3}}}
\zav{\frac{\varv_\text{K}}{100\,\text{km}\,\text{s}^{-1}}}
\zav{\frac{\dot M}{10^{-9}\,M_\odot\,\text{year}^{-1}}}}^{1/2}
\zav{\frac{R}{1\,R_\odot}}^{-1}.
\end{equation}
For a typical B star and disk parameters and $\varv_R\approx10^{-3} a$ valid for
viscosity parameter $\alpha\approx1$ follows that the strongest vertical
equatorial magnetic field that allows the instability to grow is of the order of
tens of Gauss. In the presence of turbulence, the square of the external magnetic
field $B_z^2$ can be amplified \citep{okuz,fuj}, implying the decrease of the
strongest magnetic field Eq.~\eqref{bilovices} in the disk.

\begin{figure}
\centering
\resizebox{\hsize}{!}{\includegraphics{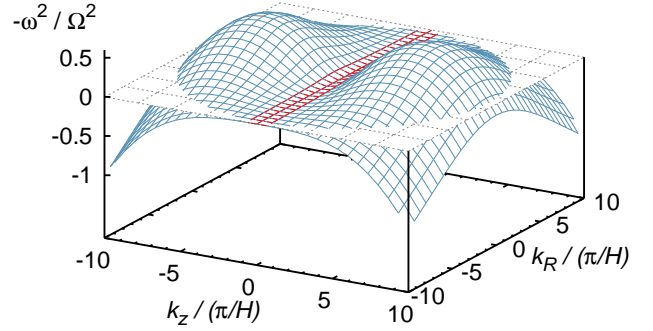}}
\caption{Zero-crossing root of the dispersion relation Eq.~\eqref{balhaw}
as a function of radial and vertical wavenumbers at $R=1.01\,\req$ for the model
given in Fig.~\ref{vd1uld}. The plane $\omega=0$ divides perturbations that are
stable $-\omega^2<0$ from that leading to an instability $-\omega^2>0$. The red
region denotes perturbations with a too large characteristic scale that can not be
accommodated in the disk $\azav{k_z/(\pi/\sqrt2H)}<1$.}
\label{omegar1}
\end{figure}

These conclusions are documented in Fig.~\ref{omegar1}, where we plot the
zero-crossing root of the dispersion relation Eq.~\eqref{balhaw} in the midplane
of the disk model described in Sect.~\ref{stacdiskkap} at $R=1.01\,\req$. This
case leads to MRI, as for some wavenumbers the root becomes negative,
$-\omega^2>0$. The perturbations with $\azav{k_z/(\pi/(\sqrt{2}H))}<1$ have
a characteristic scale larger than the disk thickness $2^{3/2}H$ and do not exist
in the disk. With increasing magnetic field intensity (or decreasing disk
midplane density) the zone of the perturbations with a too large characteristic
scale increases and fully encompasses the instability region $-\omega^2>0$ for
vertical fields stronger than that given in Eq.~\eqref{bilovice}. For such strong
magnetic fields the MRI ceases to exist close to the star. 

In stars with a magnetic field stronger than that given by Eq.~\eqref{bilovices}
the MRI does not operate in stationary conditions. Such stars may possibly
accumulate the material close to their equator until the condition for the MRI
development~Eq.~\eqref{adamov} is fulfilled. They may show episodic
ejections of matter.

Close to the star, where the parts of the Brunt-V\"ais\"al\"a
frequency $N_R$ and $N_z$ are negligible, the square of the instability branch
of MRI frequency has in the midplane from Eq.~\eqref{balhaw} an extremum for
$k_R=0$ and $k_z=\sqrt{15}\Omega/(4\varv_{\text{A}z})$, yielding 
\begin{equation}
\label{obrany}
\omega=\frac{3}{4}i\Omega.
\end{equation}
Consequently, the ratio of the MRI instability growth rate to the angular
frequency of the rotation $\omega/\Omega$ is constant close to the star. The
numerical calculations in Fig.~\ref{vd1uld} show that Eq.~\eqref{obrany} holds
nearly up to the radius, where this root of the dispersion relation changes its
sign. Moreover, Eq.~\eqref{obrany} is also valid out of the midplane, in
even slightly more radially extended region.

\section{MRI at large distances from the star}

\begin{figure}
\centering
\resizebox{\hsize}{!}{\includegraphics{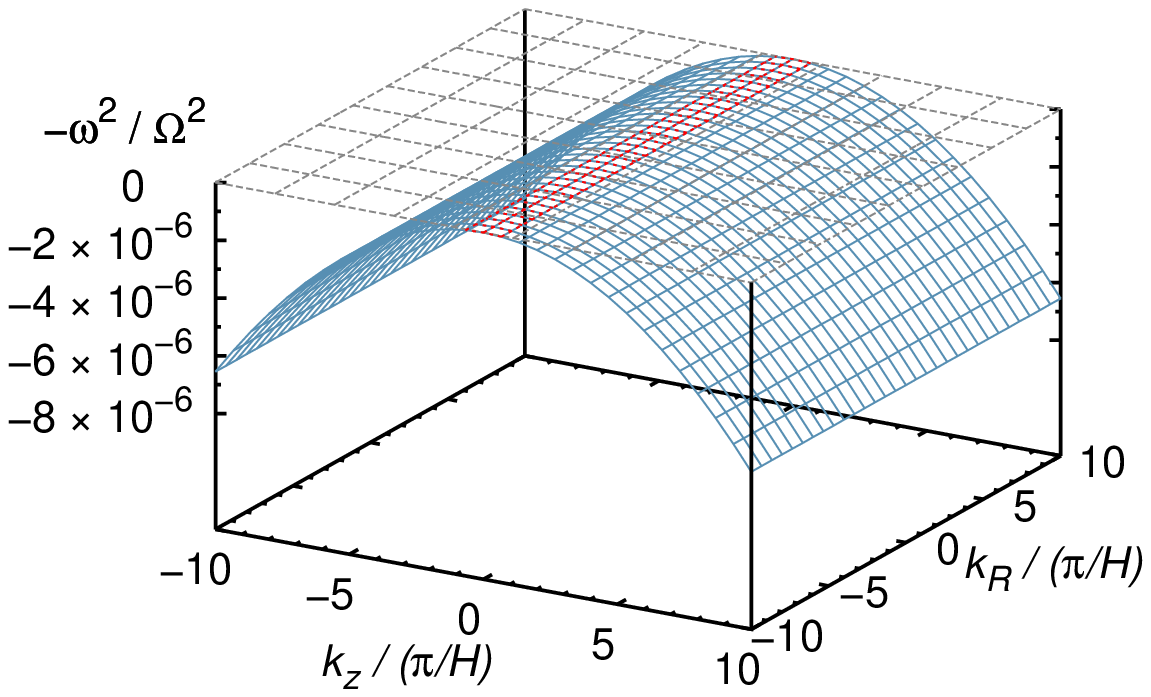}}
\caption{Same as Fig.~\ref{omegar1}, but for $R=440\,\req$.}
\label{omegar3}
\end{figure}

From the radial dependence $\varv_R/a\sim R$ and $\varv_\text{K}\sim R^{-1/2}$ the
decrease of the limiting magnetic field Eq.~\eqref{bilovices} proportional to
$R^{-1.75}$ is slower than the decrease of magnetic field intensity of the
dipolar field, which is proportional to $R^{-3}$. This shows that the disks that
develop MRI close to the star, are also unstable with respect to MRI at larger
radii. Moreover, if the magnetic field does not lead to MRI close to the
star, it may be weak enough for the development of MRI at larger radii.

However, despite this, the MRI may vanish at very large radii because of another
effect. At the locations where the epicyclic frequency becomes too low,
$\kappa^2\ll N_R^2$ the root of the dispersion relation Eq.~\ref{balhaw} that
was negative close to the star becomes positive and the instability vanishes.
This can be seen for Keplerian rotation $\kappa^2=\Omega^2$, where
Eq.~\ref{balhaw} gives in the midplane where $N_z=0$ (for concreteness for $k_R\ll k_z$) for $\kappa^2\ll
N_R^2$ the approximation for the lower root of the dispersion relation
\begin{equation}
\label{brno}
\omega^2=k_z^2 \varv_{\text{A}z}^2-\frac{4}{N_R^2}\Omega^2k_z^2
\varv_{\text{A}z}^2.
\end{equation}
With increasing radii $\omega^2$ changes its sign from negative to positive at
the point where from Eq.~\eqref{brno}
\begin{equation}
\Omega=\frac{1}{2}N_R=\frac{1}{\sqrt{10}}a
\azav{\frac{1}{\rho}\frac{\de\rho}{\de r}}.
\end{equation}
The latter identity follows from the definition of $N_R$ in Eq.~\eqref{jesenius}
assuming isothermal disks. For a radial density decrease proportional to
$\rho\sim R^{-n}$ the limiting azimuthal velocity above which the instability does
not develop is $\varv_\phi=(n/\sqrt{10})\,a\approx a$ for considered disks. The
disappearance of the MRI can be understood as a consequence of low orbital
velocity. For orbital velocities lower than the sound speed the centrifugal
acceleration becomes unimportant and the branch of the dispersion relation that
for high rotational velocities corresponds to MRI changes to ordinary Alfv\'en
waves.

This is demonstrated in Fig.~\ref{omegar3}, where we plot the root that may
cross zero of the dispersion relation Eq.~\eqref{balhaw} in the midplane
of the disk model described in Sect.~\ref{stacdiskkap} at $R=440\,\req$ (where
$\varv_\phi=a$). All roots of the dispersion relation are positive, $\omega^2>0$ and
the MRI does not occur here. We note that this occurs regardless the magnitude
of the magnetic field. The same effect also appears above the midplane, but for
slightly larger radii.

\section{Time-dependent models with decreasing viscosity}

\begin{figure}
\centering
\includegraphics[width=\hsize]{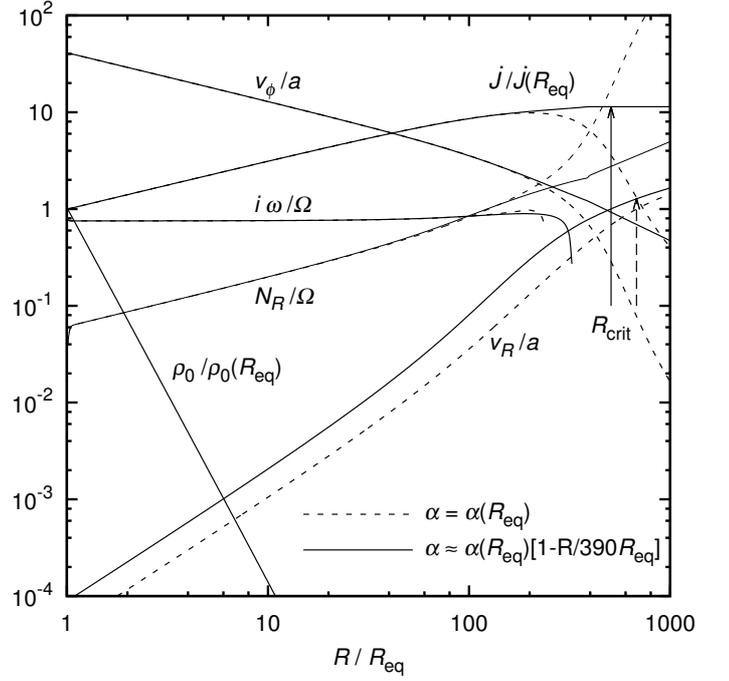}
\caption{Final stationary state of time-dependent disk models
with constant and variable viscosity parameter $\alpha$.
We plot the same variables as in the stationary model in Fig.~\ref{vd1uld}.
In the constant-viscosity model (dashed line) the viscosity parameter 
$\alpha=\alpha(R_{\text{eq}})$, while in the decreasing-viscosity model 
(solid line) the parameter
$\alpha=\alpha(R_{\text{eq}})[(391\,R_{\text{eq}}-R)/(390\,R_{\text{eq}})]$
below $R=390\,R_{\text{eq}}$ and $\alpha=\alpha(R_{\text{eq}})/390$ elsewhere.
The inner boundary viscosity parameter $\alpha(R_{\text{eq}})= 0.1$.
We consider the isothermal disk of the main-sequence B2--B3 type star 
(see Sect.~\ref{stacdiskkap}). Arrows denote the locations of the critical
points.}
\label{alp0.100}
\end{figure}

We studied the consequences of MRI disappearance at large radii using
time-dependent models with radially decreasing viscosity parameter. For these
models we used the same set of basic hydrodynamic equations, the same disk
conditions, and the same stellar parameters as in Sect.~\ref{stacdiskkap} for
B2--B3 type star. We calculated the models of isothermal disk for a constant
viscosity parameter $\alpha=\alpha(R_{\text{eq}})$ and with decreasing viscosity
$\alpha=\alpha(R_{\text{eq}})[(391\,R_{\text{eq}}-R)/(390\,R_{\text{eq}})]
\approx\alpha(R_{\text{eq}})[1-R/(390\,R_{\text{eq}})]$, that is, up to the
radius $390\,R_\text{eq}$ the viscosity $\alpha$ parameter decreases relatively
steeply and beyond this radius it is fixed at the constant value
$\alpha=\alpha(R_{\text{eq}})/390$ (see the discussion in Sect.~\ref{radproma}).
 
In the time-dependent models (unlike the stationary calculations) we can involve
the full second-order Navier-Stokes prescription of the viscous torque. This
provides physically more relevant distributions of rotational velocity (and
consequently of the specific angular momentum-loss rate) even in very distant
disk regions (for details see \citet{kurfek}). The results of the models we
obtain as a final stationary state of the converging time-dependent
calculations.

Figure \ref{alp0.100} compares the radial profiles of the relative disk midplane
density $\rho_0/\rho_0(R_{\text{eq}})$ and of the scaled radial and azimuthal
velocities $\varv_R/a$ and $\varv_{\phi}/a$ (where $a$ is the speed of sound) as
well as of the scaled specific angular momentum-loss rate
$\dot{J}/\dot{J}(R_{\text{eq}})$ (cf. Fig.~\ref{vd1uld} in
Sect.~\ref{stacdiskkap}) for the two cases of radial viscosity distribution.
The profile of the relative midplane maximum MRI growth rate $i\omega/\Omega$ is
calculated following Eq.~\eqref{balhaw} where we involve the full prescription
of the epicyclic frequency $\kappa$ (see Eq.~\eqref{epicyc}) and the radial
piece of Brunt-V\"ais\"al\"a frequency $N_R$ (Eq.~\eqref{jesenius}).

The time-dependent calculations show that the decretion disks may spread to
large radii despite the vanishing MRI instability. For a
constant viscosity distribution the rotational velocity (and and the angular
momentum-loss rate $\dot{J}$) begins to drop rapidly even below the sonic
(critical) point distance. This can be avoided in the model with a decreasing
viscosity coefficient \citep{kurfek}. Moreover, Eq.~\eqref{obrany} is also valid
for a decreasing viscosity in the Keplerian region, but the radius
where the MRI instability vanishes (the root of the dispersion relation changes
its sign) increases for a decreasing viscosity.  

From our time-dependent models \citep[see][]{kurfek} it follows that the disks
are developed to very far distances even for a radially decreasing
viscosity parameter. The reason is likely the high radial velocity in the
distant supersonic regions (even though the disk radial velocity in the distant
regions is about one order of magnitude lower than for line-driven
stellar winds). During the disk-developing phase we also recognize the wave with
supersonic speed that propagates to these far distant disk regions; we assume
that a similar transforming wave occurs and that its amplitude and velocity
depend on physical conditions (the density distribution) in the stellar
surroundings.

Our models also confirm the dependence of the sonic point distance and of the
distance of the disk outer edge (i.e., the radius where the rotational velocity
begins to rapidly decrease, see Fig.~\ref{alp0.100}) on the parameterized
viscosity and temperature profiles. The sonic point radius very weakly depends
on the viscosity profile while it is located at larger radii in the models with
steeper temperature decrease. The outer disk radius strongly depends on both
viscosity and temperature profiles and, consequently, the total angular momentum
contained in the disk and the mass of the disk increase with the decreasing
viscosity and temperature profiles \citep{kurfek}.

\section{Outer structure of Be star disks}

Some Be stars show variability, which is connected with phases of disk build-up
or dissipation. It is tempting to attribute this effect to a strong surface
magnetic field that inhibits the MRI until the material accumulates in the
equatorial plane at such high densities that the plasma $\beta$ parameter
becomes relatively large. However, such a model seems to be disfavoured by
observations that show more frequent disk variability in earlier B stars
\citep[e.g.,][]{huflo,jontys,bas} that have higher disk densities
\citep[e.g.,][]{granada}. Moreover, the rotational braking by a magnetized
stellar wind \citep{brzdud} favours low rotational velocities in magnetic
early-B stars. This seems to support the picture that Be stars only have weak
surface fields with intensities of the order of Gauss similar to those found in
some A stars \citep{vegamag,sirmag}. From Eq.~\eqref{bilovices} such weak fields
give rise to MRI in disks with mass-loss rates higher than about
$10^{-13}\,M_\odot\,\text{year}^{-1}$, which seems to be a safe lower limit for
all Be star disks.

Our results predict the disappearance of MRI at the radii of the order of
$10^2\, R_\text{eq}$ for isothermal disks (see Fig.~\ref{vd1uld}). This agrees
with the only determination of the outer disk radius from radio observations,
which is $450\, R_*$ for \object{$\psi$~Per} \citep{douta}. A similar result was
derived from optical spectroscopy by \citet{kraubedis} for a B[e] supergiant
\object{LHA 115-S 65}. On the other hand, the estimates of the disk extension
for classical Be stars from optical and near-IR regions typically give much
lower radii of the order of few stellar radii
\citep[e.g.,][]{meila,stefl,touha}. However, these estimates presumably do not
reflect the physical dimension of the disk, but just formation loci of
individual observables \citep{carpar}.

This becomes apparent when calculating the transverse optical depth, which is
given by $\tau_\nu=\int\kappa_\nu\rho\,\de z=\kappa_\nu\Sigma$, where
$\kappa_\nu$ is the opacity per unit of mass (assumed to be independent of $z$).
The disk is optically thick in the vertical direction up to radii where
$\tau_\nu=1$, which is using the continuity equation given by
\begin{equation}
\label{babice}
R=\frac{\kappa_\nu\dot M}{2\pi \varv_R}\approx 14\,{R}_\odot\,
\zav{\frac{\kappa_\nu}{1\,\text{cm}^2\,\text{g}^{-1}}}
\zav{\frac{\dot M}{10^{-10}\,{M}_\odot\,\text{year}^{-1}}} 
\zav{\frac{\varv_R}{10\,\text{m}\,\text{s}^{-1}}}^{-1}.
\end{equation}
The disk is optically thick close to the star for a given frequency, while it
becomes optically thin at larger distances, creating some kind of a
"pseudo-photosphere" \citep[e.g.,][]{kouha}. For example, for a typical disk
mass-loss rate $10^{-10}\,\text{M}_\odot\,\text{year}^{-1}$ and electron
scattering opacity with $\kappa_\nu\approx0.4\,\text{cm}^2\,\text{g}^{-1}$, the
disk is optically thick only to a few of stellar radii. A similar estimate can
be derived for hydrogen bound-free absorption, but this is complicated by the
uncertain disk ionization state.

The free-free absorption likely dominates in the radio domain at large distances from
the star. Eq.~\eqref{babice} cannot be directly used in this case because of the
dependence of the opacity on the density squared. Assuming a hydrostatic
density distribution in the vertical direction,
$\rho=\rho_0\exp\zav{-\frac{1}{2}\frac{z^2}{H^2}}$, and $ \chi_\nu
\approx\kappa_0\nu^{-3}T^{-1/2}\rho^2$ \citep{mih}, the radius at which the
vertical optical depth is unity is
\begin{equation}
\label{kotlarska}
R=\zav{\frac{\kappa_0}{8\pi}\zav{\frac{\mu m_\text{H}}{\pi k}}^{3/2}
\frac{\varv_\text{K}
(R_\text{eq})
\dot M^2R_\text{eq}^{1/2}}{(\varv_R/a)^2T^2\nu^3}}^{2/7}
\end{equation}
for the free-free absorption. Here we used the continuity equation. In scaled
quantities, Eq.~\eqref{kotlarska} reads
\begin{multline}
R=700\,R_\odot
\zav{\frac{\varv_\text{K}(R_\text{eq})}{100\,\text{km}\,\text{s}^{-1}}}^{2/7}
\zav{\frac{\dot M}{10^{-10}\,{M}_\odot\,\text{year}^{-1}}}^{4/7}
\zav{\frac{R_\text{eq}}{1\,R_\odot}}^{1/7}\\\times
\zav{\frac{\varv_R}{a}}^{-4/7}
\zav{\frac{T}{10^4\,\text{K}}}^{-4/7}
\zav{\frac{\nu}{10^{10}\,\text{Hz}}}^{-6/7}.
\end{multline}
This shows that for realistic disk parameters the extension of Be-star
disks in the radio domain is of the order of hundreds of stellar radii, and it does
not indicate the outer disk edge. Instead of this, radio observations can be
used to determine the disk mass-loss rate.

\section{Disks of Be/X-ray binaries}

\begin{figure}
\centering
\includegraphics[width=\hsize]{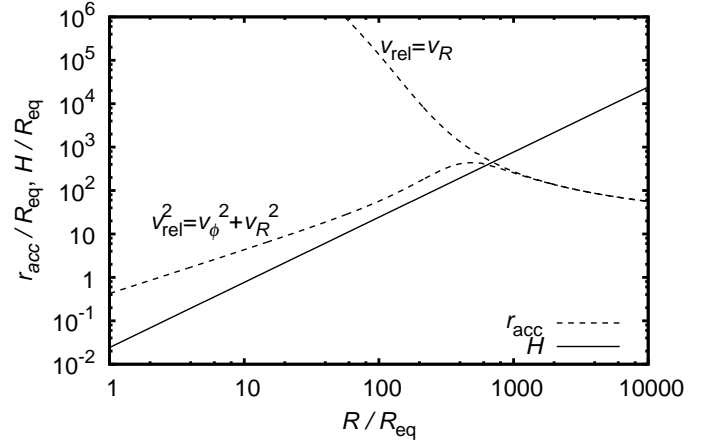}
\caption{Comparison of the neutron star accretion radius (calculated using
Eqs.~\eqref{d1} and \eqref{d2}) with the disk scale height.}
\label{t20p0D}
\end{figure}

\begin{table}[t]
\caption{Parameters of Be/X-ray binaries}
\label{bex}
\centering
\begin{tabular}{lcccccc}
\hline
Binary & Sp. Type &  $T_\text{eff}$ & $R$\tablefootmark{1} & $D$ & $\x L$ \\
&  & [kK] & [$R_\odot$] & [$R_\odot$] & [$\text{erg}\,\text{s}^{-1}$] \\
\hline
\input{bex.tex}
\hline
\end{tabular}
\tablefoot{\tablefoottext{1}{Derived from the effective temperature and spectral
type using formulas of \citet{har} and \citet{okali}.}\input{ref_bex.tex}.}
\end{table}

The X-ray emission in the Be/X-ray binaries comes from the accretion of the Be
star disk material onto the neutron star \citep[see][for a review]{reig}. The
binary separation $D$ therefore provides a constraint on the outer disk radius,
although the process of a disk truncation is complex and there may not be
a unique relationship between outer disk radius and the binary separation
\citep{okane}.

Within the classical Bondi-Hoyle-Littleton approximation, the neutron star
accretes from the radius \citep{vobe,okane}
\begin{equation}
r_\text{acc}=\frac{2G\x M}{\varv_\text{rel}^2},
\end{equation}
where \x M is the neutron star mass and $\varv_\text{rel}$ is the relative
velocity of the neutron star and the disk. There are two extreme cases for
aligned systems, either the neutron accretes material that is corrotating, in
which case
\begin{subequations}
\begin{equation}
\label{d1}
\varv_\text{rel}=\varv_R,
\end{equation}
or the disk is truncated far from the neutron star, in which case the relative
velocity may be roughly approximated by
\begin{equation}
\label{d2}
\varv_\text{rel}^2=\varv_\phi^2+\varv_R^2.
\end{equation}
\end{subequations}
If the accretion radius is larger than the disk scale height, $r_\text{acc}>H$, then the
neutron star may accrete all material from the disk, giving the X-ray
luminosity
\begin{equation}
\label{lx}
\x L=\frac{G\x M\dot M}{\x R},
\end{equation}
where \x R is the neutron star radius. In the opposite case, $r_\text{acc}<H$, only a
fraction of about $r_\text{acc}/H$ of the disk material is accreted.

In Fig.~\ref{t20p0D} we compare the accretion radius with disk scale-height for
the model displayed in Fig.~\ref{vd1uld}. For both Eqs.~\eqref{d1} and
\eqref{d2} the neutron star accretion radius is significantly larger than the
disk scale height up to the distance of the order of hundreds of stellar radii
from the Be star. We also note that the time needed to cross the distance $H$ in
the radial direction $H/\varv_R\approx({a}/{\varv_R})P_\text{orb}$ is longer
than the local orbital period $P_\text{orb}$ in the subsonic part of the disk
($\varv_R<a$). This indicates that the neutron star is able to accrete all
material from the disk if the neutron star resides in a subsonic part of the
disk. Therefore, with known X-ray luminosity one can infer from Eq.~\eqref{lx}
the disk mass-loss rate for aligned systems.

In Table~\ref{bex} we collected a sample of Be/X-ray binaries from literature.
In all cases the binary separation is relatively low, therefore the neutron star
is able to accrete all material from the disk. From Eq.~\eqref{lx} we derive for
stars in Table~\ref{bex}  an estimate of the disk mass-loss rate of the order
of $10^{-13}-10^{-9}\,\text{M}_\odot\,\text{year}^{-1}$, which quantitatively
agrees with theoretical estimates derived from evolutionary calculations of
\citet{granada}.

The temperature of the disk in Be/X-ray binaries may be affected by the X-ray
irradiation. To estimate the influence of this effect, we calculated additional
stationary models with radially increasing temperature $T=T_0(\req/R)^p$ with
$p<0$. In these models the disk critical point is located closer to the stellar
surface than in isothermal disks \citep[cf.][]{kurfek}. This means that the
neutron star is located in the region of the sonic point for many stars in
Table~\ref{bex}. For a larger binary separation the neutron star is not able to
accrete all material from the disk and the X-ray luminosity should decrease
(see~Fig.~\ref{t20p0D}). This might be one of the reasons why we do not observe
Be/X-ray binaries with very large separation.

\section{Discussion}

\subsection{Influence of the radiative heating and cooling}

The derivation of the dispersion relation Eq.~\eqref{balhaw} assumes adiabatic
perturbations. However, the outflowing disks of hot stars can be assumed to be
in the radiative equilibrium close to the star \citep{milma,carbjo}. In this
case it is more natural to assume isothermal instead of adiabatic perturbations.
This is frequently done in numerical MRI simulations
\citep[e.g.,][]{misto,papalojza,fronel}, but detailed models include
radiation transport \citep{flak}. These simulations show that MRI and a
subsequent turbulence also develops in the case of perturbations in the
radiative equilibrium. Consequently, our main conclusions remain
the same even for disks in radiative equilibrium.

\subsection{Implications for stellar mass-loss}

The disk mass-loss rate is determined by the angular-momentum loss needed to
keep the star near to the critical rotation. The larger the extension of the
disk, the larger the specific angular momentum of the disk material, and the
lower the mass-loss rate. Close to the sonic point the disk angular
momentum-loss rate flattens (see Fig.~\ref{vd1uld}) and becomes independent of
radius. MRI disappears in this region, which means that we can use the sonic
point radius to calculate the disk mass-loss rate, as was suggested by
\citet{kom}.

\subsection{Radial variations of $\alpha$ viscosity parameter}
\label{radproma}

The MRI simulations can be divided into two types. The local ones, confined to a
simulation box with a typical dimension of few scale heights
\citep[e.g.,][]{hawba,skala} and the global ones \citep[e.g.,][]{kralik,penny}.
But even the global simulations do not address the problem of the viscosity variations
at a distance of the order of hundreds of stellar radii. Consequently, one has to
rely on the analytic analysis to estimate the radial behaviour of the viscosity.
 
The order-of-magnitude viscosity estimate is $\mu\approx\rho\ell\langle
v\rangle$, where $\rho$ is the density, $\ell$ is the mean free path, and
$\langle v\rangle$ is the mean velocity. The mean free path of the perturbed
element can be estimated assuming that the radial scale of turbulent elements is
the same as the vertical one, $\ell\approx H$, and the mean velocity can be
assumed to be proportional to $\langle v\rangle\approx \azav{\omega} \ell$. This
gives for the viscosity $\mu\approx\rho\azav{\omega} H^2$, which from
Eq.~\eqref{obrany} depends on radius as $\mu\sim R^{-2}$ close to the star.
Equating this to the viscosity parameterization
$\mu\approx\alpha\rho H a$ suggested by \citet{sakura}, we conclude that $\alpha$ does not
depend on radius close to the star. This agrees with numerical simulations of
\citet{penny}. From this discussion it follows that at larger distances one can
assume $\alpha\approx \azav{\omega} R/ \varv_\text{K}$.

\subsection{Disk extension}

Our main results are valid for accretion disks as well. From
the discussion here it follows that the decretion or accretion disk extension is of
the order of the critical (sonic) point radius. For the isothermal disks
Eq.~(14) of
\citet{kom} gives
\begin{equation}
R_\text{crit}=\frac{3\,\mu m_\text{H}G}{10\,k}\frac{M}{T}\approx
10^{-5}\,\text{pc}\,\zav{\frac{M}{1\,M_\odot}}\zav{\frac{T}{10^4\,\text{K}}}^{-1}.
\end{equation}
This yields the disk extension of hundreds to thousands of solar radii for disks
around objects with mass of the order of the solar mass, while it gives tens to ten
thousands of parsec for supermassive black holes found in centres of galaxies.

\section{Conclusions}

We provided an analytic study of the MRI as a source of anomalous viscosity in
Be star disks. Our study needs to be verified by future magnetohydrodynamical simulations, but
this does not seem to alter the basic results of our paper:
\begin{itemize}
\item The MRI does not develop in stars with a strong surface magnetic field. In
these stars the plasma $\beta$ parameter is too low and the material is not
transported outwards by MRI.
\item The MRI disappears at large distances from the star, where the disk rotational
velocity is equal to the sound speed and the radial disk velocity approaches the
sound speed. In isothermal 
disks
this occurs at the radius of several hundreds of
stellar radii.
The time-dependent simulations show that the disks may disseminate to the
infinity even in the case of vanishing MRI.
\item The radial dependence of the angular momentum-loss rate flattens at the
regions where MRI disappears, consequently the relation between the mass-loss
rate and angular-momentum rate
\citep{kom}
is valid in this case.
\item We discussed the observational limits on the outer disk structure for
Be and Be/X-ray stars. For Be stars the optical and possibly
even radio observations do not trace the outer disk radius. For Be/X-ray
binaries the neutron star is able to accrete all material from the disk,
which makes the X-ray luminosity a measure of the disk mass-loss rate.
\end{itemize}

\begin{acknowledgements}
This work was supported by the grant GA\,\v{C}R  13-10589S.
The access to computing and storage facilities owned by parties and projects contributing to the 
National Grid Infrastructure MetaCentrum, provided under the program 
"Projects of Large Infrastructure for Research, Development, and Innovations" (LM2010005) is appreciated.
\end{acknowledgements}

\end{document}

%% file: bex.tex
\object{V831 Cas}\tablefootmark{ 2 }& B1V & 24 & 4.5 & 480  & $ 2\times10^{35} $ \\
\object{IGR J16393-4643}\tablefootmark{ 3 }& BV & 24 & 4.5 & 18.8  & $ 4\times10^{35} $ \\
\object{V615 Cas}\tablefootmark{ 4 }& B0Ve & 26 & 4.9 & 43  & $ 5\times10^{35} $ \\
\object{HD 259440}\tablefootmark{ 5 }& B0Vpe & 30 & 5.8 & 510  & $ 1.2\times10^{33} $ \\
\object{HD 215770}\tablefootmark{ 6 }& O9.7IIIe & 28 & 12.8 & 260  & $ 6.5\times10^{36} $ \\
\object{CPD-632495}\tablefootmark{ 7 }& B2Ve & 34 & 7.0 & 177  & $ 3.5\times10^{34} $ \\
\object{GRO J1008-57}\tablefootmark{ 8 }& B0eV & 30 & 5.8 & 390  & $ 3\times10^{37} $ \\
\object{HD 102567}\tablefootmark{ 9 }& B1Vne & 25.5 & 4.8 & 240  & $ 1.1\times10^{35} $ \\

%% file: ref_bex.tex
\tablefoottext{2}{\citet{bex1}, \citet{bex2}}
\tablefoottext{3}{\citet{bex3}, \citet{bex4}, \citet{bex5}}
\tablefoottext{4}{\citet{bex6}, \citet{bex7}, \citet{bex8}, \citet{bex18}}
\tablefoottext{5}{\citet{bex9}, \citet{bex10}}
\tablefoottext{6}{\citet{bex11}, \citet{bex12}, \citet{bex13}}
\tablefoottext{7}{\citet{hmxb21}, \citet{hmxb22}}
\tablefoottext{8}{\citet{bex16}, \citet{bex17}}
\tablefoottext{9}{\citet{bex19}, \citet{bex20}}